\documentclass[format=acmsmall, review=false, screen=true]{acmart}

\usepackage{booktabs} 
\usepackage{color}
\usepackage{multirow}
\usepackage{framed}
\definecolor{Gray}{gray}{0.7}
\definecolor{lightGray}{gray}{0.9}
\usepackage[most]{tcolorbox}
\usepackage{float}
\usepackage{xcolor,pifont}

\usepackage{soul}

\usepackage{balance}
\usepackage{flushend}
\usepackage{multicol}
\usepackage{placeins}
\usepackage{array}
\usepackage{fontawesome} 
\usepackage{microtype}
\usepackage{setspace}
\usepackage{longtable}
\usepackage{makecell}
\usepackage{pdflscape}
\usepackage{comment}
\usepackage{afterpage}

\RequirePackage{pdflscape}
\usepackage{color}
\usepackage{setspace}
\usepackage{CJK}
\usetikzlibrary{trees,positioning,shapes,shadows,arrows}

\newcommand{\find}[1]{
\begin{tcolorbox}[leftrule=1mm,toprule=0mm,bottomrule=0mm,left=1pt,right=2pt,top=2pt,bottom=2pt] 
#1
\end{tcolorbox}
} 

\tikzset{
  basic/.style  = {draw, text width=2cm, drop shadow, font=\sffamily, rectangle},
  root/.style   = {basic, rounded corners=2pt, thin, align=center, fill=white},
  level-2/.style = {basic, rounded corners=6pt, thin,align=center, fill=white, text width=3cm},
  level-3/.style = {basic, thin, align=center, fill=white, text width=1.8cm}
}
\usepackage{tcolorbox}

\newcommand{\todo}[1]{}
\renewcommand{\todo}[1]{{\color{red} TODO: {#1}}}

\usepackage{amsmath}
\usepackage{url}

\begin{document}
%

\title{On the Way to SBOMs: Investigating Design Issues and Solutions in Practice}

\author{Tingting Bi}

	\affiliation{%
		\institution{Data61, CSIRO and the University of Western Australia}
        \country{Australia}
	}
	\email{tingting.bi@data61.csrio.au}
    
 	\author{Boming Xia}
	\affiliation{%
		\institution{Data61, CSIRO and the University of New South Wales}
        \country{Australia}
	}
	\email{boming.xia@data61.csiro.au}   
    
    \author{Zhenchang Xing}
	\affiliation{%
		\institution{Data61, CSIRO and Australian National University}
        \country{Australia}
	}
	\email{zhenchang.xing@data61.csiro.au}

  \author{Qinghua Lu}
	\affiliation{%
		\institution{Data61, CSIRO and the University of New South Wales}
        \country{Australia}
	}
	\email{qinghua.Lu@data61.csiro.au}

  \author{Liming Zhu}
	\affiliation{%
		\institution{Data61, CSIRO and the University of New South Walesa}
        \country{Australia}
	}
	\email{liming.zhu@data61.csiro.au}

\begin{abstract}
The Software Bill of Materials (SBOM) has emerged as a promising solution, providing a machine-readable inventory of software components used, thus bolstering supply chain security. This paper presents an extensive study concerning the practical aspects of SBOM practice. Leveraging an analysis of 4,786 GitHub discussions from 510 SBOM-related projects, our research delineates key topics, challenges, and solutions intrinsic to the effective utilization of SBOMs. Furthermore, we shed light on commonly used tools and frameworks for generating SBOMs, exploring their respective strengths and limitations. Our findings underscore the pivotal role SBOMs play in ensuring resilient software development practices and underscore the imperative of their widespread integration to bolster supply chain security. The insights accrued from our study hold significance as valuable input for prospective research and development in this crucial domain.
\end{abstract}
\maketitle


%

\section{Introduction}

Modern software systems underpin a plethora of essential functions and services. With this heightened reliance comes an increasing need for secure and resilient software supply chains. These chains, intricate and dynamic by nature, render software susceptible to a myriad of threats. For instance, adversaries often exploit vulnerabilities within software components, leading to potentially disastrous outcomes like system compromises, data breaches, and financial losses ~\cite{ohm2020backstabber,enck2022top, ladisa2023sok}. As a response, the imperative to establish mechanisms that ensure the integrity, security, and quality of software components throughout the supply chain becomes even more pronounced~\cite{ellison2010evaluating}. Implementing these mechanisms entails integrating protocols such as code reviews, vulnerability scanning, and digital signatures, all bolstered by effective governance and risk management practices~\cite{stadtler2015supply}.

The software supply chain landscape is characterized by a blend of open-source and proprietary components, each bearing pivotal roles in system execution, development, testing, and deployment processes \cite{barclay2019towards, stadtler2015supply,enck2022top}. The comprehensive interplay of these components directly impacts software confidentiality, integrity, and availability. A disconcerting trend has emerged in the form of escalating software supply chain attacks, as evidenced by a striking average annual surge of 742\% reported between 2019 and 2022 \cite{krasner2021cost}. The repercussions of these attacks are multifaceted: malevolent software packages can jeopardize both system confidentiality and integrity \cite{schneier2019every}, while obsolete components and vulnerabilities in existing ones can disrupt system availability. This reality was illustrated when ChatGPT experienced downtime in 2023 due to a bug in the redis-py library \cite{gpt_23}.

To address these challenges, \textbf{Software Bill of Materials }(SBOM) has been identified as a critical measure for enhancing software supply chain security, as stated in a White House executive order  \cite{house_executive_2021}. The existing research has undoubtedly yielded valuable insights; however, a notable gap remains in terms of empirical evidence concerning the actual production and usage of SBOMs across the complete spectrum of the software development lifecycle. Furthermore, a scarcity of empirical investigations into discussions pertinent to SBOMs underscores the necessity of establishing optimal practices for their implementation and generation. This study endeavors to delve into the challenges surrounding SBOM usage and corresponding solutions. In doing so, it seeks to deepen our comprehension of SBOMs, thereby laying the groundwork for the assessment of best practices within the software supply chain.

Built upon our prior research \cite{xia2023empirical}, this work seeks to undertake a more comprehensive analysis of SBOM relevant discussions in real-world projects. Our primary objective is to condense the issues encountered by developers and the potential solutions pertinent to SBOMs. We examined a corpus of 4,786 SBOM relevant discussions spanning 510 projects on GitHub. Our contributions to this research are as follows:

\begin{itemize}
\item We provided the first comprehensive classification of SBOM relevant issues and the potential solutions that can be applied to address them. We also identified different phases of SBOM lifecycle and their characteristics.
\item We correlated SBOM issues with various stages in the SBOM life cycle. This analysis serves as a guide for developers, aiding in their comprehension of challenges and offering recommendations for effectively integrating SBOMs to tackle development issues in real-world scenarios.
\item We identified gaps in existing SBOM production and usage and suggested future research directions for improvements. 
\end{itemize}

The remainder of this paper is structured as follows: Section \ref{subsec_relatedwork2} describes background and related work. Section \ref{Sec_Methodology} introduces the our research design and Section \ref{sec_results} and \ref{sec_discussion} presents and discusses the results of our study, respectively. Section \ref{sec_validity} discusses the threats to validity. Finally, Section \ref{sec_conclusion} concludes this work with future directions.

\label{sec_intro}

\section{Background and Related Work }

\label{subsec_relatedwork2}

\subsection{Software Bill of Materials (SBOM)}
\label{sub_relatework_SBOM}

SBOM is an emerging research topic that has received relatively limited attention in the literature so far. We presented some previous works around SBOM in this section. For example, Carmody \textit{et al}. \cite{carmody2021building} provided a high-level overview of how SBOMs can improve the resilience of medical software supply chains. They demonstrated the benefits of SBOMs for software producers, consumers, and regulators and highlighted the progress that has been made in this area. Barclay \textit{et al}. \cite{barclay2019towards} proposed a conceptual model that combines a static bill of materials (BOM) with a dynamic bill of lots to record the static components of data as well as the dynamic data values contributing to a specific experiment. Although their work is only conceptual, it could be applied to data-driven AI systems in various fields. In 2022, Barclay \textit{et al}. \cite{barclay2022providing} introduced their work using a BOM as a verifiable credential for transparency into AI systems. More recently, Xia \textit{et al}. propose to share and exchange SBOMs via verifiable credentials on blockchain and extended the notion of SBOMs to AI bill of materials (AIBOM) ~\cite{xia2023trust}.

In addition to these contributions, a multitude of studies related to SBOMs, such as software composition analysis (SCA), release engineering, and reproducible builds, have emerged in recent years. For instance, Imtiaz \textit{et al}. \cite{10.1145/3475716.3475769} undertook a comparative analysis of various SCA tools for vulnerability reporting, while Ombredanne \textit{et al}. \cite{ombredanne2020free} conducted a comprehensive review of SCA tools for license compliance. Kengo Oka \textit{et al}. \cite{9821841} delved into the application of SCA within the automotive industry, and Mackey \textit{et al}. \cite{mackey2018building} explored the integration of automated SCA into DevOps processes.

\subsection{Mining Software Repositories}
\label{sub_relatework_MSR}

Mining software repositories is a valuable approach for understanding developers' perspectives on SBOM design issues and solutions in their projects. By analyzing data from repositories like version control systems, bug trackers, and code reviews, researchers can gain insights into how developers handle SBOM-related challenges and propose solutions. This approach has been widely used in software engineering research. For example, Rahman \textit{et al}. \cite{rahman2014insight} compared successful and unsuccessful pull requests in GitHub projects, identifying key topics and insights from the discussions. Tsay \textit{et al}. \cite{tsay2014let} investigated pull request evaluations and discussions, while Casalnuovo \textit{et al}. \cite{casalnuovo2015developer} studied collaboration dynamics within GitHub teams. Following a similar approach, we defined criteria to select GitHub projects for mining SBOM life cycle, relevant design issues, and potential solutions.

\subsection{Research Gaps}

\label{subsec_researchgap}

\begin{table*}
\caption{Comparison of the key previous works with our work}
\label{Comparison of the characteristics of related work with our work}

\centering
\scriptsize

  \begin{tabular}{c|p{2cm}|p{2cm}|p{2.5cm}|p{2.5cm}}
  
    \hline
   
 \makecell{\textbf{Work}}  & \makecell{\textbf{Methodology}} & \makecell {\textbf{Results}} & \makecell{\textbf{Software artifact}} & \makecell {\textbf{Software activity}} 
    \\
   \hline
   \hline
Hendrick \textit{et al.} \cite{lf_2022}  & A \textbf{survey} of 291 responses within Linux Foundation community  &  Summarized SBOM readiness, production, and consumption across industrial and organizations.    & Opinions and work experience from developers and organizations & Quality assurance (security and risk ).
   \\
   \hline

Balliu \textit{et al.} \cite{balliu2023challenges} & \textbf{Case Study} of six tools of SBOM generation for open source java projects.  & Focusing on comparing on the accuracy of SBOM generation for complex projects  & Java projects and six tools & SBOM documentation 

   \\
   \hline

 Shi \textit{et al.} \cite{shi2021experience}  & \textbf{Case study} of three large scale of commercial projects within Huawei company & Focusing building verifiable systems by checking SBOMs & Source code of the large-scale commercial projects and third party code.  & Software verification
   \\
   \hline
   

   Our previous work \cite{xia2023empirical}  & \textbf{Interview} $\&$ \textbf{Survey}   & Summarized three main topics and 26 statements from developers' perspective & Opinions and experience from developers  &  SBOM adoption and generation, and tooling
   \\
   \hline

   This work & Case study of mining SBOM relevant data in 510 \textbf{GitHub projects} & Classified the development issues into \textbf{three main categories} and their corresponding solutions in the real-world development contents  & 4,786 SBOM relevant discussions & SBOM documentation, 11 development activities associated with SBOM are identified\\

   \hline
\end{tabular}
\end{table*}

While the research works discussed in Section \ref{sub_relatework_SBOM} provides valuable insights into SBOM practice, there are still gaps in the research that need to be addressed to fully understand SBOM characteristics and their practical solutions by developers. This study aims to contribute to filling these gaps, and Table \ref{Comparison of the characteristics of related work with our work} compares our study's key aspects with those of other relevant studies from various perspectives:

\begin{itemize}
\item \textbf{Methodology} - Compares the methodologies applied in our work with those of previous studies.
\item \textbf{Results} - Compares the results of our research with those of previous studies.
\item \textbf{Software artifacts} - Compares the software artifacts related to SBOM analyzed in previous studies with those in our work.
\item \textbf{Software activities} - Compares the relevant software activities that can benefit from leveraging SBOM in previous studies with those in our work.
\end{itemize}

As shown in Table \ref{Comparison of the characteristics of related work with our work}, we employed the mining software repository methodology to extract comprehensive and practical data, providing insights into SBOM relevant development issues and solutions in the real-world projects. Our findings reveal the existence of four distinct SBOM development phases, encompassing 11 development activities. Furthermore, we identified three primary categories of development issues, along with their corresponding potential solutions.

\section{Research Design}
\label{Sec_Methodology}

Case studies are well-suited for gaining insights into contemporary phenomena \cite{host2012case}. Following the guidelines presented by Runeson \textit{et al.} \cite{runeson2009guidelines} and Sen \textit{et al.} \cite{sen2012open}, we conducted an exploratory study \cite{easterbrook2008selecting} aimed at investigating discussions relevant to SBOMs in GitHub projects. Our case study design is outlined as follows: in Section \ref{SBOM_discussions_GitHub}, we offered an overview of the general process that developers typically follow when engaging in SBOM relevant discussions within GitHub projects. In Section \ref{subsec_research_method_process}, we detailed our research questions and the process we employed to address them. Furthermore, in Section \ref{subsec_case_unit_of_analysis}, we described the criteria for selecting GitHub projects, as well as our data collection and analysis methods.

\subsection{SBOM Discussions in GitHub Projects}

\label{SBOM_discussions_GitHub}

\begin{figure*}
\centering

\includegraphics[width = \textwidth]{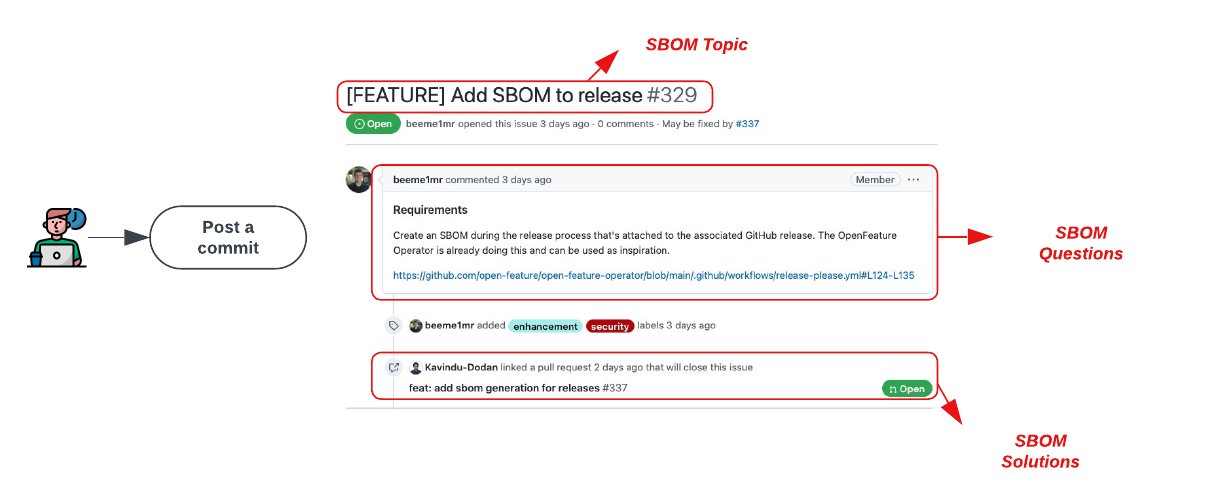}
\caption{Items for SBOM data analysis. \textbf{\textit{Annotation1}: SBOM topics}, which summarize the issues related to SBOMs; \textbf{\textit{Annotation2}: SBOM questions}, which describe the design issues developers encountered during development; and \textbf{\textit{Annotation3}: SBOM solutions}, which represent the potential solutions developers suggested for the design problems}.
\label{Fig_SBOM_issue}
\end{figure*}

As mentioned earlier, the goal of our study is to analyze SBOM relevant discussions in real-world projects, with a particular focus on GitHub, a pivotal platform for software development. We exemplified the manner in which developers discuss and respond to SBOM in Fig. \ref{Fig_SBOM_issue}. The typical discussion process involves one developer initiating the SBOM relevant topic (Annotation 1 in Fig. \ref{Fig_SBOM_issue}) along with detailed issues (Annotation 2), and other developers respond to these issues based on the given context (Annotation 3).

To ensure a comprehensive understanding of SBOMs from various perspectives, we also collected additional pertinent information for analysis, such as the time intervals between posting and closure of SBOM relevant issues. A detailed description of our data collection pertaining to SBOM discussions in GitHub projects is provided in Section \ref{subsec_case_unit_of_analysis}.

\subsection{Goal and Research Question}
\label{subsec_research_method_process}

The objective of this study is to provide a comprehensive understanding of SBOMs, encompassing different stages, characteristics, challenges, and potential solutions. To achieve this objective, we formulated three research questions (RQs). 

\textbf{RQ1. SBOM \textit{Life Cycle - How many phases that SBOMs traverse from the production to the usage in practice?}}

\textbf{Rationale}: By delineating and categorizing the SBOM life cycle, we endeavor to elucidate how many phases that SBOM go from its production to usage in practice. The insights derived from this analysis can equip researchers and practitioners with a structured and logical approach towards the challenges from the SBOM production to usage.

\textbf{RQ2. SBOM \textit{Development Issues} - What common issues are concomitant with SBOMs, and what are their characteristics, such as their open or closed status, and the duration needed for resolution?}

\textbf{Rationale}: Design issues arise when developers address SBOM concerns and identify the problem space. We explored the common design issues that developers face when producing and using SBOMs. By discerning and analyzing the inherent traits of these challenges, we can offer insights that might aid developers in forestalling or alleviating potential difficulties.

\textbf{RQ3. SBOM \textit{Development Solutions} - What solutions have developers discussed to contend with SBOM issues?}

\textbf{Rationale}: In response to development challenges, developers deliberate on design solutions and explore alternative options. Our objective, achieved via meticulous analysis of SBOM discussions, is to identify and implement effective solutions for the issues.
                                      
\subsection{Research Process}
\label{subsec_case_unit_of_analysis}

\begin{figure*}
\centering
\label{fig_overview_of_process}
\includegraphics[width = 12cm]{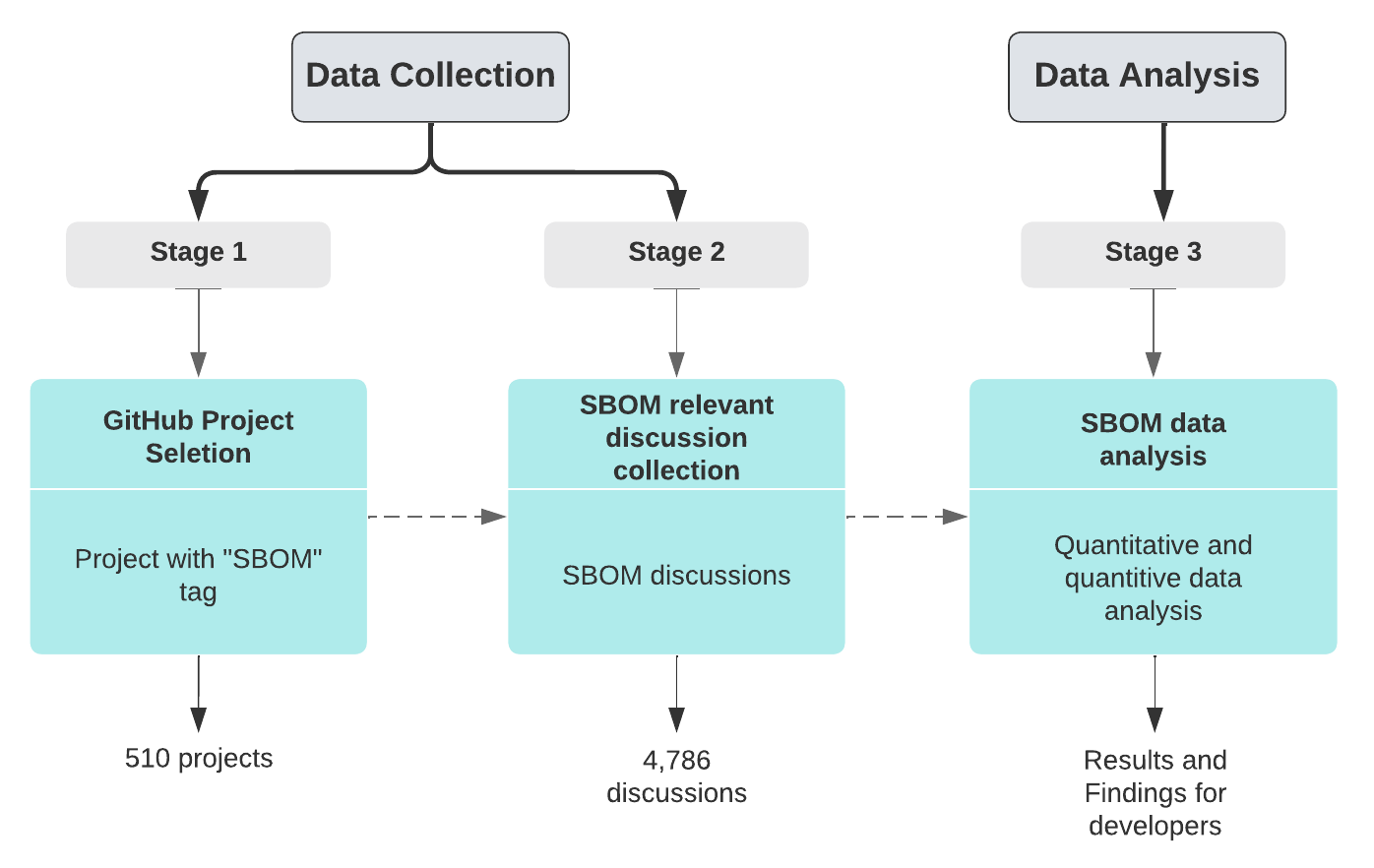}
\caption{The overview of the research methodology}
\label{Fig_Research_process}
\end{figure*}

This section outlines the process that we followed to collect and analyze SBOM discussions, as illustrated in Fig. \ref{Fig_Research_process}.

\subsubsection{Data Collection}
\label{subsubsec_data_collection}

The data collection procedure was divided into two stages, as expounded below (see Fig. \ref{Fig_Research_process}): 

\textbf{Stage 1: Project Selection - \textit{Tag-Based filtering}}. To ensure the selected repositories in this study were representative and non-trivial, we used a tag-based filtering method to choose GitHub repositories labeled with``SBOM". The primary focus was placed on the \textbf{Issue} discussions within the chosen GitHub projects. To be specific, we instituted the following selection criteria:
\begin{itemize}
    \item The repository should encompass more than 10 developers;
    \item The number of issue discussions should surpass 200.
\end{itemize}
Such criteria were established to guarantee an adequate number of developers actively participating and a rich dataset for extracting SBOM discussions. 

\textbf{Stage 2: SBOM-Relevant Data Collection - \textit{Content-Based filtering}}. In this stage, we manually collected and analyzed SBOM relevant discussions initiated by developers. Specifically, we selected issue discussions as the data resource, and through analyzing the developers' discussions, we explored common SBOM design problems and potential solutions that developers may have. The data collection process consists of two steps:
\begin{itemize}
\item Step 1: extraction of Titles, Questions, and Answers containing the keyword ``SBOM'' within the issues.
\item Step 2: compilation of pertinent data, such as project release versions, into an Excel spreadsheet for subsequent analysis.
\end{itemize}

As the results, the combined approach yielded 510 projects and 4,786 SBOM relevant discussions. We conducted the manual analysis of the discussions was favored over (semi-)automated techniques to mitigate the risk of overlooking or misclassifying SBOM discussions, which could compromise the validity of our findings.

\begin{table}
\caption{Extracted data items and their associations to RQs}
\label{table_dataiterm}
\centering
\scriptsize
\begin{tabular}{p{0.5 cm}|p{1.5 cm}|p{4cm}|p{1.5 cm} | p{2 cm}}
\hline
$\#$ & \textbf{Data Item} & \textbf{Description} & \textbf{RQ} & \textbf{Analysis Method}
\\
\hline
\hline

D1 & \textit{SBOM discussion timestamp} & This item reflects when SBOM discussions occurred & RQ1, RQ2, and RQ3 & Descriptive statistics \\
\hline

D2 & \textit{SBOM discussion status} & This item shows whether SBOM discussions are Open or Closed & RQ1 & Descriptive statistics\\
\hline

D3 & \textit{SBOM discussion content} & This item 
shows the nuances of SBOM discussions, segregating into three predefined communication types: ask, mention, and reply. & RQ2 and RQ3 & Grounded theory\\
\hline

D4 & \textit{Project releases} & This item supports the exploration of the interactions between release timelines and SBOM discussions. & RQ 1, RQ2, and RQ3 & Descriptive statistics \\
\hline

\end{tabular}
\end{table}

\subsubsection{Data Analysis}
\label{subsubsec_data_analysis}
As shown in Fig. \ref{Fig_Research_process}, we conducted quantitative and qualitative data analysis. 

\textbf{Stage 3: SBOM data analysis}. The correlation between the extracted data items and their respective analysis methodologies is shown in Table \ref{table_dataiterm}. The qualitative (D1, D2, and D4) and quantitative (D3) data extracted from the discussions were evaluated per the research questions delineated in Section \ref{subsec_research_method_process}.

To analyze the qualitative data (e.g., D3, the content of SBOM discussions), we employed a bottom-up and systematic encoding approach from Grounded Theory \cite{strauss1997grounded}. Bottom-up approaches are suitable for classifying specific domain knowledge and concepts when there is no predefined and existing concepts in that domain \cite{de2014exploratory}. Note that, in our analysis, only open coding and axial coding techniques from Grounded theory were employed. The process comprises three steps which were executed iteratively: 
\begin{enumerate}
    \item \textit{Open coding}, executed by first two authors, split the SBOM discussions in the commits into separate parts, i,e., words, phrases, or sentences, which were labeled as concepts;
    \item \textit{Axial coding}, executed by first two authors and confirmed with another author, was employed to identify categories (i.e., the main topics and subtopics) through refining and relating the concepts generated in open coding to a category;
    \item \textit{Verification}, to reduce personal bias in coding, any inconsistencies on the coding results are cross examined by the all the authors. 
\end{enumerate}

Descriptive statistics, such as means, standard deviations, and frequency distributions, are used to analyze the quantitative data (D1, D2, and D4) in SBOM discussions, providing insights into the characteristics of SBOMs. 

\section{Results}

\label{sec_results}

\subsection{RQ1: SBOM Life Cycle}
\label{subsec_results_RQ1}

\begin{figure}
\centering
\includegraphics[width = 10 cm]{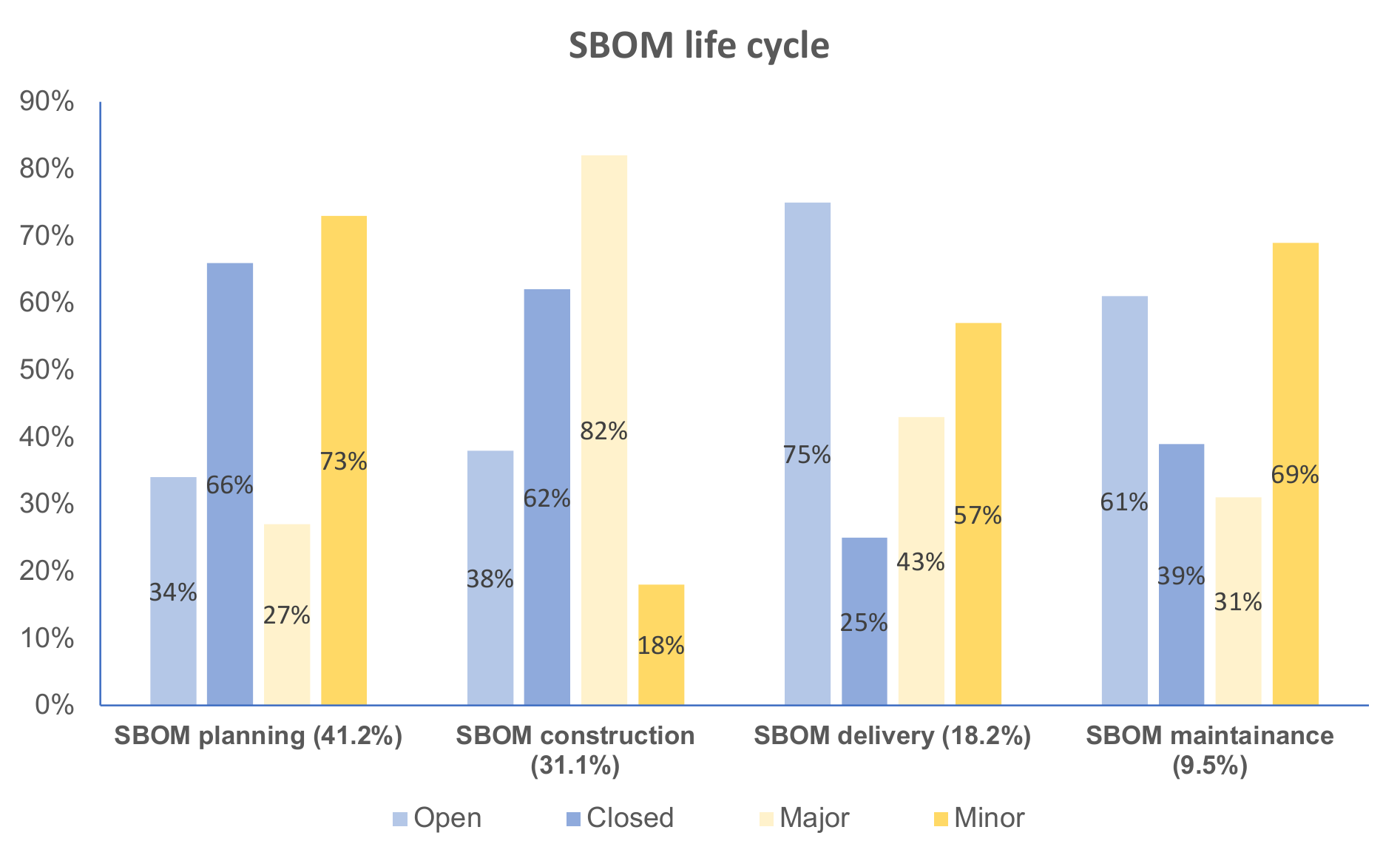}
\caption{Four phases of SBOM life cycle and their status and releases}
\label{Fig_Development_phases}
\end{figure}

To address RQ1, we analyzed the tags, titles, contents of SBOM discussions, and the releases of the projects, specifically data items D1, D2, D3, and D4 in Table \ref{table_dataiterm}. The aim was to identify how many stages in SBOM life cycle and their characteristics. As a result, we classified the collected discussions into four distinct phases of SBOM life cycle, which are: \textit{SBOM planning}, \textit{SBOM construction}, \textit{SBOM delivery}, and \textit{SBOM maintenance}. Fig. \ref{Fig_Development_phases} illustrates the four phases of SBOM life cycle, along with the statistical information on their corresponding project releases and the issue status.

\textbf{1. SBOM planning phase} (41.2$\%$). During the SBOM planning phase, developers typically discuss the dependencies and relationships among software components and their associated licenses. This helps them identify potential non-functional requirements, compliance issues, and supply chain risks \cite{mohammed2017exploring,longshore2022managing}. They also discuss procedures for updating and maintaining the SBOM throughout the software development life cycle, including integration with version control and automated build systems. Furthermore, developers discuss how to share and communicate the SBOM with other stakeholders, such as customers, regulators, and third-party vendors, is an essential aspect of SBOM planning. This involves selecting a suitable format, such as SPDX or CycloneDX, and defining policies and procedures for access control, data privacy, and confidentiality. We identified several activities in planning phase as below: 

\begin{itemize}

\item  \textbf{Non-functional requirement identification}: The Non-functional requirement of a software product is significantly impacted by the quality of the development processes that support it. Non-functional requirement assessment is, therefore, a crucial planning activity. During the SBOM planning phase, developers discuss non-functional requirement and how SBOMs impact it. For example, developers plan to generate SBOMs that enhance specific non-functional requirements, such as transparency and cyber security (we delved into it further in Section \ref{SBOM_issue_clssification}). 
\item \textbf{Risk identification}: Effective risk management is crucial for the successful planning and execution of a project. Within the context of SBOM relevant risk discussions, developers discussing the impact of using third-party packages, share expertise, and collaborate to minimize potential risks.
\item  \textbf{Preparing SBOM documentation}: Developers discuss the lack of standardization or adopting standards when planning and documenting SBOMs. There are no universal standards for documenting SBOMs, and the existing standards, such as SPDX, CycloneDX, and SWID tagging, are inconsistent and confusing. Developers discuss a set of issues involving this activity, highlighting the need for standardization in SBOM documentation to ensure consistency and accuracy across different software products and organizations.
\item \textbf{Integrating compliance}: integrating compliance into SBOM planning is crucial to ensure that software development process aligns with regulatory and organizational requirements. For example, some organizations may need to identify open source licenses or comply with export control regulations \cite{nadgowda2022engram}. These discussions underscore the importance of flexibility in SBOM construction and delivery to meet the varying compliance needs of different organizations.
\end{itemize}

In addition, to analyze the relationships between the project releases and SBOM planning, we analyzed the releases of projects. Our results show that the majority of SBOM planning discussions (i.e., around 73$\%$) take place during minor releases, and around 66$\%$ of these discussions (e.g., SBOM planning for security concerns) remain in an open status (see Fig. \ref{Fig_Development_phases}). The discussions of SBOM planning in major releases can facilitate understanding the changes of software components and dependencies included in the release, thereby enabling the tracking of the root cause of any development issues. Some examples of SBOM planning discussions are as follows:

\vspace{0.1 cm}

\faCommenting  \ ``\textit{AMI is \textbf{planning} \textbf{to roll out an entire ecosystem for firmware SBoM}, likely based on CycloneDX. CycloneDX is discussing evidence inclusion for the upcoming spec release in another firmware area, uswid is being suggested and prototyped with, though lacking proves/references; it's rather decoupled from the actual binaries it describes there is an org named Veraison dealing with verification and attestation at large, mostly rooted in TCG DICE, including SWID, which is an ISO NISTIR thing SBOMs-Data, Column: 5 | Row: 94)}"

\faCommenting  ``\textit{Describe the \textbf{\textit{development plan}} you've considered. A clear and concise description of the plan to make the solution ready. It can include a development timeline, resource estimation, and other related things. Integrate Anchore Syft into Harbor and just like you did for Claire, update the UI to account for this and then allow Harbor the option of checking\textit{\textbf{ SBOMs against vulnerability scans}} to determine Low Medium, or High Threshold for said container or Helm Chart}."

\vspace{0.1 cm}

\find{

\textbf{Summary 1} -  Developers \textbf{\textit{plan SBOMs by identifying dependencies of components, updating and maintaining procedures}}, and discussing non-functional requirements, risks, standardization, and compliance along it. In addition, most SBOM planning discussions occur during minor releases. 
}

\textbf{2. SBOM construction phase} (31.1$\%$).
SBOM construction phase involves the discussions on data extraction of software components, including their names, versions, and licenses, as well as any dependencies and relationships among them. Once this information is collected, it is organized and formatted according to the chosen SBOM standard. The resulting SBOM also include additional metadata, such as descriptions, copyright notices, and security and vulnerability information \cite{ding2019method}. In addition, during the construction phase, developers verify the completeness, accuracy, and consistency of the generated SBOM by comparing it with other sources, such as version control systems, issue trackers, and package managers. The activities involved in the SBOM construction phase include SBOM generation and re-generation, as well as quality assessment or verification. We identified three main activities below:

\begin{itemize}
\item \textbf{SBOM generation}: The SBOM generation activity involves creating a nested inventory that lists the ingredients comprising software components \cite{ding2019method}. This involves outlining or updating the information and processes necessary to support the fundamental and essential features of SBOMs.   
\item \textbf{SBOM re-generation}: As software evolves over time, new components are added, existing components are updated, and dependencies between components may change, all of which can have an impact on the overall security and integrity of the software. It is important to create and maintain an accurate and updated SBOM to reflect changes in software composition. SBOMs need to be re-generated upon software compositions changes to stay up-to-date. Having an (re-generated) SBOM can aid in change management efforts by providing a clear understanding of what has changed between different versions of a software product.
\item \textbf{SBOM quality assessment and verification}: SBOM quality assessment involves verifying whether the SBOM is complete, accurate, and up-to-date. The high-quality SBOM should enable management of vulnerabilities and configuration, and incident response. In addition, developers compare the generated SBOM to other sources of information, such as vulnerability databases or licensing information, to ensure its accuracy and completeness. By engaging in these discussions and taking these steps, developers can ensure that the SBOM is comprehensive and accurate, ultimately improving the overall security and compliance of the software product.

\end{itemize}

The SBOM construction phase is critical for maintaining the integrity and reliability of software products and services. By creating and verifying a comprehensive and accurate inventory of software components and their attributes, developers can better manage the software supply chain, identify and mitigate potential risks, and improve the overall quality and security of software products and services. The SBOM construction phase consists of generating or re-generating SBOMs (upon systems changes) and the verification of generated SBOMs. Developers also establish procedures for managing and updating the SBOMs, including version control policies, workflows for reviewing and approving changes, and tools for automatically updating the SBOMs when new software components are added or changed. Three activities involved in the SBOM construction phase are SBOM generation, SBOM re-generation, and SBOM quality assessment.

It should be noted that an SBOM can not really be ``updated"; rather, it should be re-generated with a new timestamp \cite{xia2023empirical}.  However, it is common for developers to use the term ``update". Our analysis shows that around 82\% of SBOM constructions occur during major releases, and 62\% of SBOM discussions have been resolved and closed. When an SBOM regeneration is discovered in a minor release, having an SBOM can facilitate identification of the components requiring updates or patches. Moreover, SBOM regeneration in minor releases primarily aims to provide a list of all software components and their dependencies included in a particular minor version of the software product.

\faCommenting\ ``\textbf{\textit{Generating SBOM describing the source in the Cilium repository}} \textit{using bom takes, on average, 10 minutes. As a result, the CI build time increases by 30 minutes if we generate an SBOM describing the source for all three CI images in Image CI Build and the CI ultimately fails, throwing an error that no space is left on the runner.(SBOMs-Data, Column: 4 | Row: 114)}"

\faCommenting\ ``\textit{We have the fix for Issue $\#$134 but we are still seeing concurrent issues when running the SBOM tool. \textbf{Is it possible to update the code to have a more unique value in it? Like, a process id or a guid?} Please let me know if I can provide any more details. (20. Microsoft/Component detection, Pos. 6-8)}"

\faCommenting  ``\textit{Propogate image-resolve-mode for \textbf{SBOM generator image} $\#$3446 I discovered that updates to th}e: edge tag for buildkit-syft-scanner weren't re-running the scanner, even though we had a new image."

\find{
\textbf{Summary 2} -  The SBOM construction phase entails extracting and organizing data on software components, creating and verifying a comprehensive and accurate inventory, all with the aim of managing the software supply chain, mitigating risks, and improving software quality. \textbf{\textit{SBOM generation, re-generation, and quality assessment and verification are crucial activities during this phase}}.
}

\textbf{3. SBOM delivery phase} (18.2$\%$). Developers discuss to deliver SBOMs to the appropriate stakeholders in a timely and secure manner. This phase involves defining distribution and delivery mechanisms, such as email, web portals, or APIs, and considering the format to meet the specific needs of stakeholders. Ensuring the security and privacy of delivering SBOMs is also crucial, developers discuss potential risks and threats such as data breaches or unauthorized access. In addition, encryption, access controls, or other security measures are also discussed by developers to deliver SBOMs \cite{mohammed2017exploring}. We identified one major activity in SBOMs delivery phase:

 \begin{itemize}

     \item Secure delivery:  The discussions involve secure SBOM file transferring/distributing protocols and secure storage to prevent unauthorized access and tampering. 
\end{itemize}

Furthermore, we found that around 57$\%$ of SBOM delivery discussions took place in the \textit{minor} releases and around 75$\%$ of discussions remain \textit{open} status (see Fig. \ref{Fig_Development_phases}). Minor releases introduce small changes or bug fixes to existing products, as such, it reflects that it is important to ensure SBOMs for minor releases accurately record these changes and updates. Some example discussions as below:

\faCommenting  ``\textit{Upload a BOM file generated by cyclonedx/cyclonedx-npm (not sure if special conditions apply here regarding the packages, we identified at least 2 projects BOMs showing this behaviour) \textbf{Upload a very different BOM file to update the Dependency-Track content}. In our case we replaced the NPM tool in one project with the webpack-cyclonedx tool, which produced a completely different BOM (as expected).}"

\faCommenting\ ``\textit{Demo-for-Microsoft-SBOM-Tool. This is a comparison of Microsoft's SBOM Tool against a CycloneDX BOM Generator to show how Microsoft's tool detects transient dependencies.}"

\faCommenting\ ``\textit{Figure out how to automate the \textbf{SBOM creation} -> a user cloning our template should get this action automatically configured and enabled Propagate this change to our existing policies. In a quick research I found out that the current tool used to generate the SBOM files for Rust and Go policies support Swift. But in my quick try, the tool failed. Thus, this issue also include a research if the tool in use really works and how to use it}."

\find{
\textbf{Summary 3} -  Developers discuss and define delivery mechanisms and security measures to ensure timely and \textbf{secure distribution} of SBOMs in delivery phase. 57$\%$ of SBOM delivery discussions occur in minor releases, with 75$\%$remaining unsolved.
}

\textbf{4. SBOM maintenance phase} (9.5$\%$). This phase encompasses the establishment and enforcement of policies and procedures aimed at ensuring the proper governance of SBOMs \cite{chulani2008software}. It involves defining rules and standards governing the generation, storage, and sharing of SBOMs across stakeholders. Effective data governance is crucial to the successful integration of SBOMs, as it ensures the accuracy, consistency, and security of SBOMs throughout their life cycle. Our analysis revealed that 69$\%$ of SBOM governance discussions occur during major releases. We identified two key activities of SBOM governance and provided their corresponding examples:

\begin{itemize}

\item SBOM data maintenance: Discussions revolve around the difficulty of tracking and managing all the software components and dependencies in an application, particularly when they are constantly changing, even if there is an SBOM in place.
\item Integration of SBOMs into existing processes: Developers discuss how integrating SBOM maintenance with existing development, testing, and deployment processes can pose a challenge, particularly if those processes lack documentation or standardization.
\end{itemize}

\faCommenting  ``\textit{SBOMs should form part of your vulnerability \textbf{maintainance process} by using them to scan for vulnerabilities when acquiring software from the supply chain and also understanding your vulnerability posture when releasing software to your users. As vulnerabilities are being discovered continuously, vulnerability scanning of released software should be proactively performed so that your users can be informed of any new vulnerabilities as they are discovered}."

\faCommenting  ``\textit{A SBOM is a nested inventory, a list of ingredients that make up software components. \textbf{The SBOM maintenance work has advanced since 2018 as a collaborative community effort}, driven by National Telecommunications and Information Administration’s (NTIA) multistakeholder process}."

\find{

\textbf{Summary 4} -  In SBOM maintenance phase, discussions are about to \textbf{\textit{ensure accuracy, consistency, and security}}. Discussions mostly occur in major releases, with two activities identified: data management, and integrating SBOMs into existing processes.
}

\subsection{RQ2: SBOM Development Issues}
\label{subsec_results_RQ2}

We answered RQ2 from two perspectives: first, we analyzed the D3 and D4 data items (see Table \ref{table_dataiterm}) to classify the collected SBOM discussions into categories (as explained in Section \ref{SBOM_issue_clssification}). Then, we analyzed the D1 and D3 data items to gain insights into the time duration required to resolve each category of SBOM issue (as described in Section \ref{Howlong_issuesolved}).  

\subsubsection{SBOM Development Issues}
\label{SBOM_issue_clssification}

\definecolor{cellorange}{rgb}{ 1,  .949,  .8}
\definecolor{cellgreen}{rgb}{ .776,  .878,  .706}
\definecolor{cellblue}{rgb}{ .608,  .761,  .902}
\definecolor{celllightgrey}{rgb}{ .921,  .921,  .921}
\definecolor{celldarkgrey}{rgb}{ .664,  .664,  .664}

\definecolor{shadecolor}{rgb}{.92,  .92, .92} 

\afterpage{%
\begin{landscape}

\begin{table*}[p]
\tiny
\caption{SBOM issue categories}
\label{table_discussed_issue}
\centering 
  \begin{tabular}{ p{2cm}| p{3 cm}|p{4cm}|p{9cm} |p{0.8cm}}
    \hline
\textbf{Main Category} &  \textbf{Sub Category} & \textbf{Description}  & \textbf{Example}  & \textbf{\makecell{\textbf{$\%$}}}  
    \\
   \hline
   \hline

\multirow{2}{2cm}{\textbf{SBOM internal quality issue}} & 

\textbf{N/A} & This category discusses the quality, for example, the documentation and inaccuracy and incomplete of SBOMs. & Example 1: ``\textit{The following differences are detected: Package version changes, Package licence changes, Package removed, Package added}"

\textit{Example 2: ``With regard to barriers to adoption from a CycloneDX perspective. The availability of quality tooling that produces SBOMs is not an issue. The ease of doing so is not an issue. The CycloneDX Tool Center has ~150 known tools documented, with many commercial and open source projects having adopted it and provided the capability to their users."}

\
& \makecell{15.3$\%$} \\


\hline

\multirow{2}{2cm}{\textbf{SBOM relevant development issue}} &   \textbf{Traceability issue} & This category discusses the issues that contain the details and supply chain relationships of various components used in building the software. & \textit{Example 1}: ``\textit{By uploading the latest bill-of-materials from every CI build, the BOM-Bar service can keep an inventory of \textbf{all packages} in use within an organization and provide feedback to the development team on potential (license) policy violations}." 

Example 2 ``\textit{These are quite hard to trace back to the original distro package that they were copied from, and as a result, no tool (that I'm aware of) will mark the corresponding distro \textbf{package} in the SBOM}." 

\textit{Example 3:} ``\textit{Select a preferred SBOM format (likely cyclone-dx) generate SBOMs during assemble and package, investigate options to validate, merge existing SBOMs (inputs $\&$dependencies). Investigate options to combine, translate SBOM formats into a single one.Open questions:do other \textbf{packagers} require SBOM support?} " 

& 

\makecell{55.2$\%$}   \\

&  \textbf{Dependency issue} & This category discusses the vulnerable dependency issues by leveraging SBOMs  & \textit{Example 1:} ``\textit{\textbf{Dependency Security Scanner} that automatically notifies you about vulnerabilities}."  

\textit{Example 2: ``Dependencies in CycloneDX SBOM format not parsed $\#$ 206 While trying guac with a CycloneDX SBOM I found that no edges are being created. To make sure this is a reproducable case you could run the following commands to create the same SBOM: It seems no edges are identified in that BOM and therefore no \textbf{dependency} graph is built in neo4j, only single nodes are created. But as far as I understand the SBOM actually contains the dependency information which could be used to build the graph.''}

& \makecell{45.2$\%$}  \\

& \textbf{Non-functional requirement issue} &  This category discusses the security and vulnerability issues that could casue damage to software or the information on them, as well as the services other systems or organzations provide. & \textit{Example 1}: ``\textit{After feeding an SBOM generated by Anchore/Syft, I discovered that I got no \textbf{vulnerabilities} back, when I was expecting multiple.}"

\textit{Example 2: ``\textbf{Incompatibility} with SBOM from @cyclonedx/cyclonedx-npm $\#$2265 We found some kind of incompatibility with some SBOMs generated by @cyclonedx/cyclonedx-npm. We could not reproduce this with SBOMs from the older NPM tool or other sbom tools."}

& \makecell{25.1$\%$} \\

& \textbf{License compliance issue} & This category discusses issues about software products comply with open source licenses and other legal requirements. &  \textit{Example 1}:``SBOM In Cyclonedx format not providing license information as per standard"

\textit{Example 2: ``Audit \textbf{licensing and copyright entries} in the packages $\#$340 In the SBOM you can add a custom license ref, but we're not sure how to handle it in the apk, which will then flow back to the SBOM but without the context."}
& \makecell{10.3$\%$} \\

\hline

\multirow{2}{2cm}{\textbf{SBOM tooling issue}} & \textbf{Insufficient tool support} & This category is about the discussions on implementation or applying SBOM tools (i.e., automatic documentation tool), especially for complex software products.  & \textit{Example 1:} ``\textit{When generating first-party SBOMs, it's hard to know if you're generating something good (e.g. rich metadata that you can query later) or not. This tool hopes to quantify what a well-generated SBOM looks like}.'' 

Example 2: ``\textit{Support Generating SBOM Files $\#$939 You all do a wonderful job identifying licenses and comparing them against a per-configured policy. ... \textbf{There are some very good tools in the SBOM generation space}, but your ability to detect licenses is far superior to theirs. Additionally, they do not offer an ability to enforce a license policy as you have}."

& \makecell{9.2$\%$}\\

& \textbf{Automatic SBOM generation} & This subcategory is about SBOM relevant technical issues (i.e., coding issues). & \textit{Example 1: }``\textit{I have followed the below steps: Download sbom-spdx-generator binary for the windows Run sbom-spdx-generator -f json
Observe the SBOM file in spdx format. (3. 12 No usable version of libssl was found on Linux $\#$ 43, Pos. 7-11)}" 

\textit{Example 2: ``Develop the Q2 / R3 interaction pattern for having an OpenC2 Producer shepherd the process of getting SBOMs into the data store, and informing the decision maker where to get the information. Q2 / R2 is similar, but if we want the \textbf{decision maker} to operate with read-only access to the data store, it seems better to have the OpenC2 Producer churn through the URLs and populate the data store so the decision maker's \textbf{interface} is consistent regardless of where the SBOMs actually originated.}"

&  \makecell{8.1$\%$} 

\\

& \textbf{SBOM tooling quality} & This subcategory is about discussing the quality of tool to generate SBOMs. & Example: ``\textit{The availability of \textbf{quality tooling} that produces SBOMs is an issue. The ease of doing so is not an issue. The CycloneDX Tool Center has ~150 known tools documented, with many commercial and open source projects having adopted it and provided the capability to their users.} " &  \makecell{5.2$\%$} \\

\hline

\end{tabular}
\end{table*}

\end{landscape}
}

Software development involves complex issues that must be addressed efficiently and accurately for successful product delivery \cite{jain2015systematic, ragunath2010evolving}. We analyzed the D3 and D4 data items (see Table \ref{table_dataiterm}) and classified the 4,786 SBOM discussions (i.e., commits, feature requests, and bug reports) into a fine-grained taxonomy of SBOM issues. 

Our classification identified three main categories of SBOM issues: \textit{\textbf{SBOM internal quality issue}}, \textbf{\textit{SBOM relevant development issue}}, and \textbf{\textit{SBOM tooling issue}} (shown in Table \ref{table_discussed_issue}).

\textbf{1. SBOM internal quality issue}, this category refers to issues that are inherent to the SBOM itself, such as incomplete or inaccurate documentation. This category of SBOM issues accounts for 15.3$\%$, including issues related to incomplete or inaccurate information, missing data, incorrect formatting, and automatic documentation errors.

\textbf{2. SBOM relevant development issues}, 
this category pertains to discussions regarding development issues and how SBOMs can be applied for improving such issues, such as security and transparency concerns of the systems. This category can be further divided into four subcategories:

\begin{itemize}
\item \textbf{Transparency issue}: accounting for 55.3$\%$ of the discussions, this subcategory encompasses system-internal traceability problems, such as those between packages, and system-external ones (e.g., library and API usage). The aim of addressing these issues is to enhance the identification of vulnerable software components that contribute to cybersecurity incidents. Developers are discussing how to leverage SBOMs to improve traceability and establish software supply chain transparency, ultimately playing a crucial role in bolstering the trustworthiness of software systems.
\item \textbf{Dependency issue}: discussions in this subcategory revolve the complex third-party dependencies of systems. Developers raise concerns regarding change management issues, given that software systems are typically maintained by multiple developers and undergo a significant volume of code and components changes on a daily basis. As such, developers discuss the importance of leveraging SBOMs to maintain a comprehensive list of such dependencies and their respective releases. \cite{brundage2020toward}.
\item \textbf{Non-functional requirement issue}: this subcategory (accounting for 25$\%$) focuses on ensuring or improving the system's non-functional requirement issues, and developers discuss that SBOMs can be applied for addressing those issues. These non-functional requirements include classic and emerging ones (i.e., security).
\item \textbf{License compliance issue}: this subcategory (accounting for 10.3$\%$) refers to license compliance issues identified and extracted in SBOMs, including license conflicts and the lack of license tracking.
\end{itemize}

\textbf{3. SBOM tooling issues}: this category discusses the usage of SBOM relevant tools that is further classified into three subcategories: 

\begin{itemize}
\item Insufficient tool support: accounting for 9.2$\%$ of the discussions, this subcategory encompasses topics such as searching for SBOM tools, comparing relevant SBOM tools, and discussing other tools for different purposes.
\item Automatic SBOM generation: approximately 8.1$\%$ of the discussions focused on SBOM generation tools, highlighting potential challenges in capturing complex dependencies between different components, particularly in cases where multiple layers of dependencies exist or where dependencies are difficult to trace.
\item SBOM tooling quality: this subcategory accounts for 5.2$\%$ of the discussions and highlights the importance of tooling quality to ensure the accurate and secure generation, management, and sharing SBOMs. Poor tooling quality can result in errors, inconsistencies, and security vulnerabilities in SBOMs. 
\end{itemize}

\find{\textbf{Summary 5 - SBOM issue category}. Three main categories of SBOM issues that are discussed by developers. These issues are prevalent throughout the \textbf{\textit{entire development life cycle}}, and it is crucial to address them effectively to ensure the successful production and usage of SBOMs.}

\subsubsection{How Long Does It Take to Fix Various SBOM Issues?}
\label{Howlong_issuesolved}

This section aims to shed light on the relationships between various SBOM issues with issue resolution time. We measured issue resolution time as the number of days between issue reporting and closing. Table \ref{table_howlong_tofixissue} shows the minimum, maximum, mean, and median issue resolution times for various SBOM issue categories.

We found that the minimum time required to resolve all SBOM issue categories was less than a day. Most of these issues were reported by project developers who had already identified possible solutions, or had obtained information from external sources. However, we also found that the maximum issue resolution time could take several months or even years, with Automatic SBOM Generation taking the longest time to resolve (398.3 days).

Our findings highlight the importance of addressing SBOM issues promptly, as some issues may require significant time and resources to resolve. The results suggest that developers should be proactive in identifying and reporting SBOM issues to avoid delays in resolution. The results show that the resolution times of various SBOM issue categories, which can be used to inform and improve the SBOM design and development process.


\find{\textbf{Summary 6 - SBOM issue resolution time:} On average, SBOM issues take around 25 days  to be resolved. However, our analysis revealed that 65.3$\%$ of SBOM issues remain unresolved, indicating a significant need for continued efforts to address SBOM-related challenges.}

\begin{table}
\caption{Duration of SBOM issues resolution in days}
\label{table_howlong_tofixissue}
\centering 
\footnotesize
  \begin{tabular}{p{4 cm}|p{1 cm}|p{1cm}|p{1 cm}|p{1cm}}
    \hline
\textbf{Issue Type} & \textbf{Min} & \textbf{Max} & \textbf{Mean} & \textbf{Median}
    \\
   \hline
   \hline

\textbf{Traceability issue}  & 0.5 &  192.8 & 36.6 &  16.2 \\
\hline
\textbf{Insufficient tool support}  & 0.9 &  321.4 & 35.8 &  20.2 \\
\hline
\textbf{Dependency issue } & 0.4  &  321.4 & 26.6 &  29.2 \\
\hline
\textbf{Non-functional requirement issue}  & 0.4 &  432.5 & 24.9 &  45.2 \\
\hline
\textbf{License compliance issue} & 0.3 &  213.3 & 22.1 &  21.2 \\
\hline
\textbf{Automating SBOM generation} & 0.5 &  398.3 & 20.1 &  43.2 \\
\hline
\textbf{SBOM tooling quality} & 0.6 &  187.3 & 18.1 &  25.2 \\
\hline
\textbf{SBOM quality issue} & 0.7 &  153.7 & 9.1 &  38.6 \\
\hline

 \end{tabular}

\end{table}

\subsection{RQ3: SBOM Development Solutions}
\label{subsec_results_RQ3}

\definecolor{cellorange}{rgb}{ 1,  .949,  .8}
\definecolor{cellgreen}{rgb}{ .776,  .878,  .706}
\definecolor{cellblue}{rgb}{ .608,  .761,  .902}
\definecolor{celllightgrey}{rgb}{ .921,  .921,  .921}
\definecolor{celldarkgrey}{rgb}{ .664,  .664,  .664}

\definecolor{shadecolor}{rgb}{.92,  .92, .92} 


\begin{table*}
\scriptsize
\caption{SBOM Potential Solutions}
\label{table_solution}
\centering 
  \begin{tabular}{ p{2 cm}| p{3 cm}|p{3cm}|p{4.5 cm}}
    \hline
\textbf{Phase} &  \textbf{Category} & \textbf{Design Problem}  & \textbf{Solution Strategy}  
    \\
   \hline
   \hline

\multirow{2}{2cm}{\textbf{SBOM internal quality issue}}
 & 

\multirow{2}{2cm}{N/A} &  Incomplete or inaccurate SBOM documentation  & 

1. Manually update SBOMs

2. Use alternate sources

3. Analyze the the components and dependencies of software

4. Use software composition analysis tools 

\\

& 

 &  Formatting inconsistencies or lack of standardization &

1. Use a standardized format

2. Collaborate with suppliers

3. Implement continuous monitoring

\\

\hline

\multirow{2}{2cm}{}
 & 

\multirow{2}{2cm}{Traceability issues} &  Failure capture changes  &

1. Applying SBOMs to implement a robust change management process;

2. Combing SBOMs and automated tools for tracking changes;

3. Leveraging SBOMs to establish clear ownership and responsibility.

\\


& Dependency issues &  inaccurate version information 

& 

1. Applying SBOMs to identify all dependencies;

2. Verify version information in SBOMs;

3. Tracking information in SBOMs to check for vulnerabilities;

4. Applying SBOMs as a kind of data resource to ensure consistency between releases;

5. Leveraging SBOMs to monitor changes between releases.

\\


\textbf{Solutions for SBOM relevant development issue} & Non functional requirement issues discussed by developers & Non-functional requirement issues& 

1. Security: combing SBOMs to address security issues, for example, to conduct a security assessment of the software and identify potential vulnerabilities in SBOMs; 

2. Reliability: leveraging SBOMs to identify potential failure points in the software and conduct regular testing, for example, SBOM documents dependencies, developers can ensure that the updated library does not introduce compatibility issues or unintended consequences;

3. Maintainability: to document the software's architecture, design, and code. This information can be included in the SBOM and used to track changes and updates to the software over time;

4. Efficiency:  developers can quickly identify the components documented in SBOMs of the application and assess the potential risk. 

 \\

&  License compliance issues & Software is used or distributed in ways and by cross-referencing the SBOM with a database of known licenses, the organtinaztions can proactively identify any potential license violations. & 1. Applying SBOMs to identify the non-compliant software component; 

2. Negotiate with stakeholders;

3. By tracking SBOMs, developers can obtain a different license;

4. Update the license compliance in SBOMs;

5. Applying SBOMs to maintain good standing within the open-source community.
\\
\hline

\textbf{SBOM tooling issue} & Insufficent tool support issues & Insufficient tool usage can be challenging for developers to implement SBOM as part of their software development and management process.

& 1. Alternating manual process for documenting SBOMs;

2. Developer custom tooling for SBOMs;

3. Collaborated with suppliers for SBOMs;

4. Advocate for industrial SBOM standards;

5. Leverage thrid-party solutions for SBOMs production.

\\

 & Automatic SBOM generation issues & There exists inaccurate and up-to-date SBOMs that available for all the components used in their projects. & 1. Integrate SBOM generation into the development process. 

2. Use automated scanning tools

3. Utilize machine learning algorithms

4. Implement continuous monitoring

5. Leverage third-party solutions \\
\hline

\end{tabular}
\end{table*}



\label{subsec_RQ3_potentialSolution} 

In this section, we presented the potential solutions for SBOM relevant issues discussed by developers. Table \ref{table_solution} provides an overview of the general solutions categorized by SBOM issues classified in Section \ref{SBOM_issue_clssification}. The columns ``Phase" and ``Category" align with our taxonomy discussed in RQ1. Additionally, the ``Design Problem" column corresponds to the issues of topics identified in RQ2 (refer to Table \ref{table_discussed_issue}), and the last column, ``Solution Strategy," briefly describes the general solution strategies. Overall, we identified 33 high-level solutions for the SBOM relevant issues in the taxonomy. We provide a detailed explanation of each solution below:

\textbf{\textit{1. Solutions to SBOM internal quality issues}}. We identified four general solution strategies and their detailed descriptions for SBOM internal quality issues that are listed in Table \ref{table_solution}.  We found that SBOM internal quality issue (e.g., incomplete documentation or incorrect information of SBOM) varies in different cases, and most questions need to solved case by case. Generally, developers discuss four strategies: (1)  update the SBOM consistently, the most straightforward solution discussed by developers to update the SBOM with the missing information. This may involve contacting the software vendors or suppliers to obtain the necessary information, reviewing the source code, or using automated tools to generate an SBOM; (2) developers discuss to apply alternate sources; for example, consulting the National Vulnerability Database (NVD), the Common Vulnerabilities and Exposures (CVE) database, or third-party vulnerability scanners to identify known vulnerabilities in the software to improve the suitable of inaccurate or incomplete SBOMs; (3) developers discuss to analyze the software, which can help identify missing components, dependencies, and other critical information required for an SBOM. This analysis can be manual or automated, depending on the software complexity and available tools; (4) at last, developers discuss using the tools, and some analysis tools can automatically identify and track software components and their dependencies, including open source libraries and third-party code. These tools can also generate an SBOM that provides a comprehensive list of all software components and their versions used in the software.

In addition, developers propose various solutions to tackle SBOM standard  issues, for example, inconsistent formatting issues across different releases. To address these issues, we identified three general solution strategies and along with their detailed descriptions (see Table \ref{table_solution}): (1) developers suggest adopting a standardized format, such as SPDX or CycloneDX, which can help ensure consistency in the SBOM outputs. These formats provide clear guidelines on how to structure the data, making it easier to compare and analyze different software components; (2) collaborating with suppliers to gather accurate and complete data about the software components can help improve the quality of the SBOM and comprehensive outputs; (3) SBOM outputs should be continuously monitored and updated as new information becomes available. Implementing a continuous monitoring system can help ensure that the SBOM output formatting remains accurate and up-to-date.

\find{\textbf{Summary 7 - Solutions for SBOM internal quality issue:} Developers apply four strategies to solve SBOM internal quality issues, such as updating consistently, applying alternate sources, and analyzing the software. To tackle standard issues, they suggest adopting several standardized formats, collaborating with suppliers, and continuously monitoring and updating SBOM outputs.}

\textbf{\textit{2. Solutions for SBOM relevant development issues}}

\textbf{SBOM solution for traceability issues}. Developers discuss three general solutions in terms of traceability issues that identified in SBOMs: (1) a change management process can help ensure that all changes to the software components are tracked and documented. This includes changes to dependencies, versions, and other attributes that impact SBOMs. A change management process or tool can help ensure that the SBOMs is updated when new vulnerabilities or security issues are discovered; (2) applying tools to solve the traceability issues; for example, software composition analysis (SCA) or supply chain security tools can help track changes to the software components automatically. These tools can also provide alerts when new vulnerabilities are discovered in the software components of SBOMs; (3) ensuring the clear ownership and responsibility for the SBOM can help maintain its accuracy and timeliness. Establishing well-defined roles and responsibilities for managing the SBOMs can facilitate timely resolution of traceability issues.

\textbf{SBOM solutions for dependency issues}. We summarized five general solutions in term of dependency issues identified in SBOMs: (1) it is important to identify all dependencies of the software components and document them in SBOMs, both direct and indirect. Direct dependencies are explicitly declared by the software component, while indirect dependencies are required by the direct dependencies; (2) verifying the version release information for each dependency and ensure that it is correctly included in the SBOMs; (3) employing the SBOMs to check for known vulnerabilities in the dependencies, utilizing vulnerability databases or other sources of information. If any vulnerabilities are found, take appropriate actions to mitigate them; (4) checking that the version release information in the SBOMs matches that in the software component and resolve any discrepancies to ensure accuracy; (5) implementing a process to monitor changes to the dependencies and update the SBOMs accordingly. This includes monitoring for new vulnerabilities or changes to the version information.

\textbf{SBOM solutions for non-functional requirement issues}. In terms of non-functional requirement issues identified in SBOMs, developers discuss the different solutions for different quality attributes.

\begin{itemize}
    \item Security issues: developers discuss that: (1) to address security issues is to conduct a security assessment by applying SBOMs of the software and identify potential vulnerabilities; (2) additionally, creating a list of security requirements that the software must meet. This can be included in the SBOMs and used to track compliance with security standards.
    \item Reliability: some potential solutions are identifying potential failure points in the software and conduct regular testing to ensure that these points are addressed. This information in SBOMs can be used to track reliability over time.
    \item Maintainability: it is important to document the software architecture and code, and the information included in the SBOMs can be used to track changes and updates to the software over time.
    \item Efficiency: to address efficiency issues that identified in SBOMs, it is important to conduct performance testing and identify potential bottlenecks in the software, and this information in SBOMs can be helpful to track performance over time.
\end{itemize}

\textbf{SBOM solutions for license compliance issues}. Developers discuss five solutions, which can address license compliance issues in SBOMs. (1) for a non-compliant software component is not critical to the product's functionality, it can be replaced with a compliant alternative in SBOMs; (2) for the non-compliant component is a third-party one, document a compliant license or to replace the non-compliant component with a compliant one in SBOMs; (3) for the software component is open source, obtaining a different license that is more compatible with the project's needs and document it in SBOMs; (4) updating the license compliance process if compliance issues are identified in SBOMs to prevent similar issues from occurring in the future; (5) when the license compliance issue is complex, it is necessary to seek piratical advice to determine the best course of action and document it in SBOMs.

\find{\textbf{Summary 8 - SBOM solutions for development relevant issues:} Developers suggest 17 different SBOM solutions for traceability, dependency, non-functional requirement issues, including change management, tool application, and clear ownership.}

\textbf{Solutions for tool usage issue}. While insufficient tool support for SBOMs can be challenge, there are five general solutions developers discussed that can help organizations produce accurate and up-to-date SBOMs: (1) while manual processes can be time-consuming and error-prone, it is necessary if tool support for generating SBOMs is limited. Manual processes can include reviewing software licenses and component versions and manually compiling a list of all software components and their associated licenses; (2) developers discuss that if the requirements of the projects requires to develop their own tools to generate SBOMs. This may require the assistance of software developers or third-party vendors with expertise in SBOMs and tool development; (3) developers discuss that collaborating with software vendors and suppliers to ensure the SBOM is as part of their products; (4) developers discuss that organizations need to adopt industrial standards for SBOMs, which can help to promote the development of more tools and solutions that support SBOMs; (5) developers discuss to apply third-party solutions, which can help generate SBOMs. These solutions can be used to supplement existing tools or to provide a complete solution for generating SBOMs.

\textbf{Solutions for automatic SBOM generation issues}. Developers discuss five solutions that can address automatic SBOM generation issues (1) developers suggest to integrate SBOM generation into the software development process, and organizations can automatically generate SBOMs as part of their software build process. This can be done by using build tools that support SBOM generation, such as Maven, Gradle, or npm; (2) developers discuss that there are several automatic tools available that can be used to identify software components to generate SBOMs, and by integrating these tools into the software development process, SBOMs can be generated automatically for each software build; (3) developers discuss that machine learning algorithms can be trained to recognize software components and their associated licenses. This can help automate the SBOM generation process and improve the accuracy of the generated SBOMs; (4) continuous monitoring can be used to ensure that SBOMs remain accurate and up-to-date over time, and this can be done by automatically scanning software components for new versions or updates and generating updated SBOMs as necessary; (5) developers suggest several third-party solutions that can facilitate automatic SBOM generation, such as software composition analysis (SCA) tools.

\find{\textbf{Summary 9 - Solutions for SBOM tooling issues:} Developers suggest combing manual processes, developing own tooling, and collaborating with vendors, industry standards, and third-party for solving to SBOM relevant tooling issues.
}

\section{Discussion}

\label{sec_discussion}

\begin{figure*}
\centering
\includegraphics[width = 14.5cm]{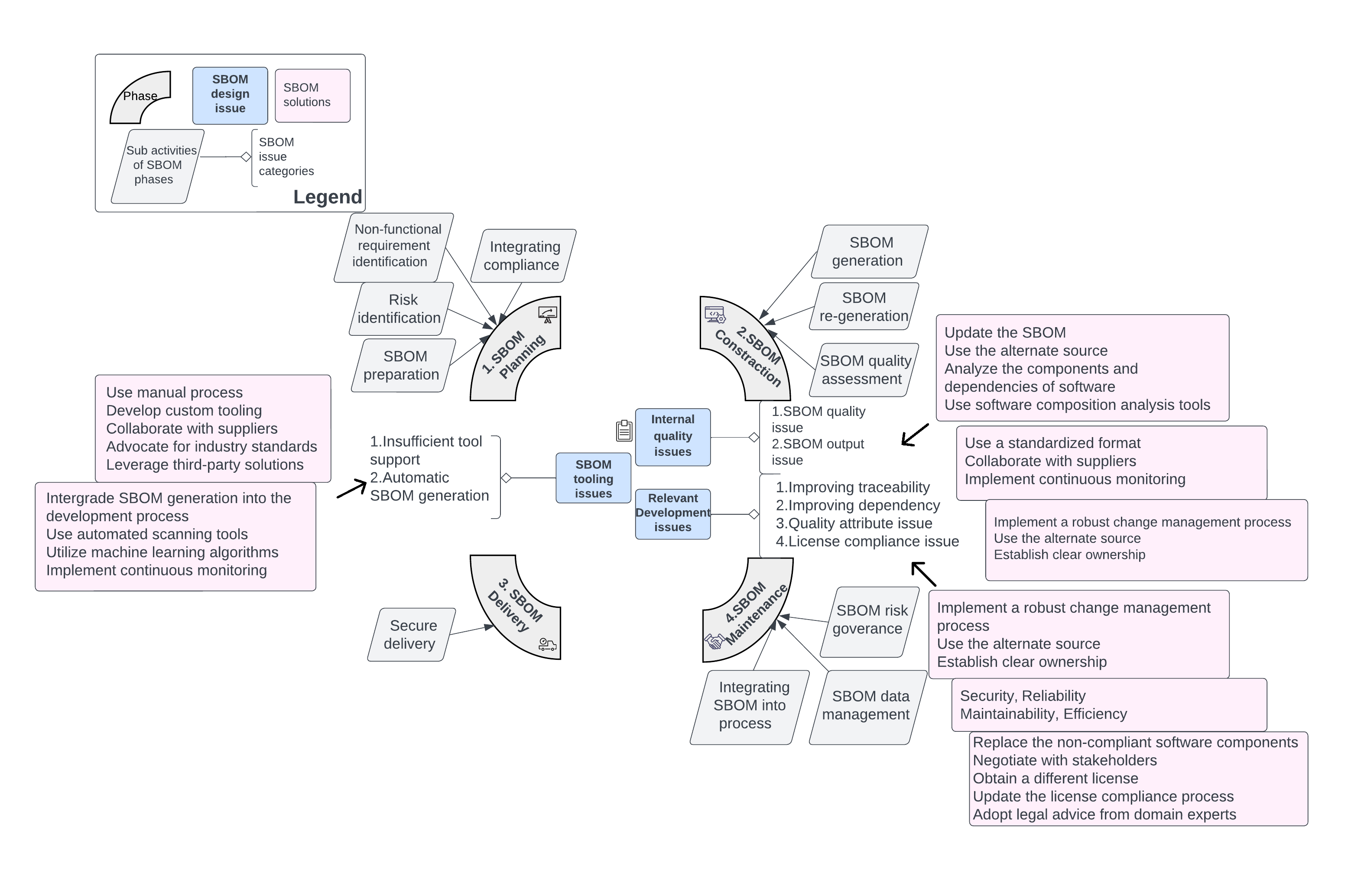}
\caption{The summary of results}
\label{Fig_results_summary}
\end{figure*}

In this section, we presented the actionable discussions from the results of three RQs (see Fig. \ref{Fig_results_summary}).

\textbf{SBOM and development activities}. Based on the reults of RQ1 (refer to Section \ref{subsec_results_RQ1}), we identified four phases in SBOM life cycle and their activities (i.e., SBOM \textbf{\textit{planning}}, \textbf{\textit{construction}}, \textit{\textbf{delivery}}, and \textbf{\textit{maintenance}}) that span across the entire software development life cycle. As such, it is crucial to maintain an up-to-date SBOM and track relevant development activities (i.e., software maintenance) to ensure the security and stability of software systems. In addition to evaluating a project's maturity and community support level, an SBOM can also help in identifying and addressing potential vulnerabilities and dependencies. It is important to apply the same level of version control and roll-back capability to data stored in SBOMs, as any changes to data can significantly impact the overall functionality and security of  software systems. 

\textbf{SBOM and technical debts}. Base on the results of RQ2.2 (see Section \ref{Howlong_issuesolved}), in average, it takes 20.3 days to fix SBOM issues on GitHub, and some of SBOM issues (e.g., non-functional requirement issues) remain unsolved and without follow-up discussions. Given the analysis of this data item, some SBOM issues could be forming technical debts. Potential reasons to solve SBOM issuse is hard could be the lack of standards or guidelines for SBOM production, leading to inconsistencies in the way SBOMs are production across different projects, and this further result in an SBOM that is difficult to maintain, understand, or update, creating technical debts. It is therefore essential to identify technical debts in SBOMs as software systems evolve over time, with new components being added and old ones being deprecated, resulting in changes to the SBOM. By applying SBOM accordingly, technical debts can be identified and avoided in the early stages of software development.

\textbf{SBOM and transparency}. Base on the results of RQ2 (see Table \ref{table_discussed_issue}), there is a set of development issues (i.e, dependency issues) identified in SBOM discussions. By integrating SBOMs, developers can have a better understanding on the software and how it is developed. For example, by providing a detailed list of software components, an SBOM can improve communication between software developers, vendors, and customers, and this can help ensure that all parties have a common understanding of the software components and dependencies. Additionally, SBOMs can increase visibility into the software supply chain by tracking the origin and provenance of each component. This is also beneficial for identifying potential security or licensing risks associated with the use of specific components. SBOMs also helps to enhanced trust that can build trust between stakeholders. Furthermore, having an SBOM can help stakeholders make informed decisions about the software product; for example, users can better assess the security and compliance risks associated with using the software.

\textbf{SBOM and traceability}. Based on the results of RQ2 (see Table \ref{table_discussed_issue}), 55.2$\%$ of traceability issues identified in SBOM discussions.  Developers need to manage large amounts of data from different resources, so it is critical to allow the developers to manage the data in a machine-readable format for efficiency; for example, SBOMs can provide tracking information that relates each component of systems. Furthermore, the primary purpose of SBOMs is to uniquely and unambiguously track components and their relationships to one another. 

\textbf{SBOM and non-functional requirements}. Based on the results of RQ2 (see Table \ref{table_discussed_issue}), 25.1$\%$ of non-functional requirements identified in SBOM discussions. To be specific, developers discuss that SBOMs can effectively improve non-functional requirements. For example, SBOMs can help developers reduce the risk and cost of software development and maintenance. For example, SBOMs can help developers to identify and address security vulnerabilities, reduce the risk of supply chain attacks, and ensure compliance with security standards and regulations. In addition, SBOMs can improve the maintainability of software by providing a comprehensive list of software components and dependencies, which can help developers identify and update outdated or vulnerable components, track the use of open-source software, and manage the software supply chain more effectively. 

\textbf{SBOM and project management}. Based on the results of RQ1 (see Fig. \ref{Fig_Development_phases}), there is a close relationships between SBOMs and projects releases. As such, SBOMs can also be used as an effective tool for project management. To ensure that SBOMs are effective for project management, it is essential to keep them concise and straightforward. The more complicated the SBOM, the more challenging it will be to integrate and adopt across teams. Additionally, any information that is likely to change or evolve over time, such as version numbers or software updates, should be separately managed. Furthermore, SBOMs should be linked to other project management tools, such as issue tracking systems and project planning tools. Such linking can help to identify dependencies and enable developers to track progress accurately. By integrating SBOMs into the project management process, developers can identify and prioritize tasks more efficiently.

\textbf{SBOM and tooling support}. Base on the results of RQ2 and RQ3, developers discuss two categories of SBOM tooling issues and a set of solutions (see Table \ref{table_solution}). The successful implementation of SBOMs requires the use of appropriate tooling to manage and maintain the data. There are many different software tools available for SBOM production and management, and one of the essential features of SBOM tooling is the ability to automate the generation of SBOMs. Automated SBOM generation can streamline the process and ensure that the information is accurate and up to date. Another critical feature of SBOM tooling is the ability to integrate with other software tools used in the development process. For example, integration with issue tracking systems (i.e., version control) can help to identify and track vulnerabilities and technical debt. Integration with project management tools can help to prioritize tasks and allocate resources effectively. Furthermore, the tooling for automatic SBOM generation tooling should be supported by facilitated by incorporating established standards. These standards should be designed to accommodate future research efforts aimed at refining and enhancing the SBOM generation process.

\textbf{Data-BOM/AIBOM}. 
Incorporating AI components into a system introduces a unique set of security and vulnerability challenges, such as data poisoning, model replication, evasion, and exploitation of traditional software flaws \cite{grotto2021vulnerability}. To mitigate these risks, it is crucial to have an accurate and up-to-date SBOM that includes all the components, including data and models, in an AI system (i.e., Data-BOM or AIBOM).
Our research findings from RQ2 indicate that developers recognize the importance of applying SBOM to AI systems. Similar to traditional software systems, the assembly of AI systems typically involves using open-source and commercial components that may contain vulnerabilities or security flaws. An SBOM can aid in identifying these components, ensuring that they are up-to-date, and responding promptly to new vulnerabilities. Moreover, an SBOM can track the provenance of AI models and datasets employed in the AI system\footnote{\url{https://www.cisa.gov/news-events/news/software-must-be-secure-design-and-artificial-intelligence-no-exception}}. This information is critical in maintaining transparency, accountability, and explainability in AI systems. By providing a comprehensive view of the system's components and their relationships, Data-BOMs/AIBOMs can improve the security, reliability, and trustworthiness of AI systems.

\section{Threats to Validity}

\label{sec_validity}

\textbf{External Validity} is about how much our results can be generalized. Whilst we mined GitHub projects extensively, it does not mean that GitHub projects contain all the SBOMs discussions that are known to the SE community. As such, the results may not be applicable to other projects that have different development purposes. We acknowledged that whilst our evidence appears to be consistent with the observations of previous works, our claims are limited to our data set. The use of this systematic approach based on encoding within Grounded Theory to analyze developer concerns and discussion can be employed to analyze communication data from other sources (i.e., Stack Overflow). 

\textbf{Construct Validity} is the extent to a measure can account for the theoretical structure and characteristics of measure. There are two threats to the construct validity of this study: (1) the first one in our study is whether the SBOM issues were extracted correctly, and in particular, the search term used to collect related issues in GitHub projects. This is a keyword-based search, and one possible threat to construct validity is that the results got from the search the search terms we used may not cover all the relevant SBOM issues. We did a pilot search to prove that most of SBOM issues were labelled by ``SBOM" tag. In addition, through applying a systematic approach to search the relevant issues, we partially reduced the degree of this threat. (2) The second one is the manually data labelling and analysis. To reduce this threat, we conducted a pilot data labelling and analysis of 100 SBOM issues before the formal data labelling, and we formulated the inclusion and exclusion criteria about whether a candidate issues should be included or not. To alleviate the threat of data extraction, the first author rechecked the data extraction results after the data extraction was conducted jointly with the second author, and any disagreements were discussed and resolved with the third author. To mitigate the threat of data analysis, the first author continuously discussed with the second author during the analysis process to reach an agreement. 

\textbf{Reliability} refers to whether the study gets the same results when other researchers replicate it. The threats to the reliability of this study concern the processes of SBOM discussion collection and analysis. We made explicit the process of how to collect and analyze data in this study. We also employed a systematic encoding approach to manually analzye the qualitative data in this work, to partially improve the reliability of the analysis results. In addition, we provided the study data set containing all the extracted data and labelling results of SBOM issues for validation. These measures partially mitigated the threats.

\section{Conclusion}

\label{sec_conclusion}

In this study, we analyzed the SBOM relevant design issues and solutions, extracting 4,786 discussion from 510 projects. Our findings show on how SBOMs are used in practice and highlighted the importance of secure SBOM in practice, developing SBOM tools, relating SBOMs with other artifacts and activities, and ensuring transparency and traceability.

As the use of SBOMs continues to gain momentum, there are several areas for future research. One important area is the development of secure SBOM development practices to ensure the accuracy and completeness of the data. Another area is the development of SBOM tools that automate the generation and maintenance of SBOMs and integrate with other development tools. In addition, relating SBOMs with other artifacts and activities is also an essential area for future research. As software development becomes increasingly complex, it is essential to understand how SBOMs relate to other artifacts, such as issue tracking systems and project management tools, and how they can be used to improve development processes. Finally, transparency and documentation are critical for the effective use of SBOMs. As we noted, the SBOM concept could be applied to AI-based products and systems, starting with the provenance of the training data, the designation of the algorithm used, how the system was tested, and other key disclosures. Future research should focus on developing best practices for documenting SBOMs and making it transparent and accessible to all stakeholders.





%

\bibliographystyle{ACM-Reference-Format}
\bibliography{reference.bib}

\end{document}